\documentclass{aa}  

\usepackage{graphicx}
\usepackage{txfonts}
\usepackage{subcaption}         
\usepackage{natbib}                                
\usepackage{lscape}             
\usepackage{amsmath}            
\usepackage{setspace}
\usepackage{placeins}           
\usepackage[inline]{enumitem}
\usepackage{xcolor}
\newcommand{\formaldehyde}{$\mathrm{H_2CO}$}

\newcommand{\methanol}{$\mathrm{CH_3OH}$}

\newcommand{\gstate}{$ 1_{10}-1_{11}$}
\newcommand{\fstate}{$ 2_{11}-2_{12}$}

\newcommand{\zerostate}{$ 1_{11}$}

\newcommand{\taug}{$\tau_{4.8}$}
\newcommand{\tauf}{$\tau_{14.5}$}

\newcommand{\nhtwo}{$n_{\mathrm{H_2}}$}
\newcommand{\tk}{$T_k$}

\newcommand{\xhtwoco}{$X_{\mathrm{H_2CO}}$}
                                
\begin{document}

   \title{Notes on the formaldehyde masers}

   \author{D.J. van der Walt
        }

   \institute{Centre for Space Research, North-West University, Potchefstroom, 2520, South
     Africa\\
             \email{johan.vanderwalt@nwu.ac.za}
   }
   \date{Received September 30, 20XX}

   \abstract
         {The 4.8 GHz formaldehyde masers are rare when compared, for example,
         to the class II methanol masers, especially when both masers are associated with
         high-mass star-forming regions. Apart from the rarity of the masers, which has
         not yet been explained, the non-detection of associated 14.5 GHz masers is also
         still an outstanding question.}  
       { The first aim of the present work is to investigate, from a theoretical point of
         view and with more realistic free-free spectral energy distributions (SEDs), whether
         there are physical conditions in the molecular envelope under which the \gstate{}
         transition is inverted but not the \fstate{} transition. The possibility that the
         non-detection of 14.5 GHz masers is due to the masing region being projected
         towards the edge of a background hyper-compact \ion{H}{II} region is also
         investigated. Since the \gstate{} transition of ortho-\formaldehyde{} is known to
         have an anti-inversion behaviour for typical conditions associated with high-mass
         star-forming regions, it is possible that attenuation affects the 4.8 and 14.5
         GHz masers. The second aim is to estimate to what extent attenuation of the 4.8
         and 14.5 GHz \formaldehyde{} masers in the molecular envelope can explain the
         small number of detected \formaldehyde{}
         masers.  }
       { The photo-ionisation code Cloudy was used to calculate more realistic free-free
         SEDs for a given spectral type of the ionising star and
         different radial dependences of the initial \ion{H}{I} distribution. The
         free-free SED obtained from the Cloudy simulation was
         used as the pumping radiation field in the statistical equilibrium calculations.}
       { Using a fit from the Cloudy simulations to the observed free-free SED of the
         hyper-compact \ion{H}{II} region G24.78+0.08 A1, it is found 
         that while the \gstate{} transition is weakly inverted, the \fstate{} transition
         is not inverted. In this case, inversion of the \gstate{} transition is dominated
         by collisions and the contribution of the free-free radiation field to the
         inversion is negligible. Analysis of the dependence of the inversion of the
         \gstate{} and \fstate{} transitions on distance into the molecular cloud
         suggests that there are regions in the circumstellar envelope where the \gstate{}
         transition is inverted but not the \fstate{} transition. The optical depths at
         4.8 and 14.5 GHz were calculated for three different dependences of the
         abundance of o-\formaldehyde{} on depth into the molecular cloud, which shows
         that significant attenuation of the maser emission is possible.  }
       { Not all hyper-compact \ion{H}{II} regions have free-free SEDs that are able to
         produce strong enough 4.8 GHz 
         masers. Attenuation of the 4.8 GHz maser emission in the molecular envelope can
         be so significant that the 4.8 GHz maser emission is completely absorbed.  Detection of
         the 14.5 GHz maser associated with the 4.8 GHz maser is not a requirement to
         prove the free-free pumping of the 4.8 GHz masers.}


   \maketitle

\nolinenumbers
\section{Introduction}
The 4.8 GHz ortho-formaldehyde (henceforth referred to as \formaldehyde{}) masers are
classified as rare, with only 22 individual masers associated with 19 high-mass
star-forming regions detected in the Galaxy. The scarcity of these masers is in stark
contrast with the significantly larger number of class II methanol, water, and OH masers
associated with high-mass star-forming regions. The fact that ortho- and
para-\formaldehyde{} are widely detected in the Galaxy and the use of these molecules as
probes of the physical conditions in molecular clouds \citep[see e.g.][and references
  therein]{Downes1980,Mangum1993, Tang2018a, Tang2018b, Brunthaler2021, Mahmut2024}
further highlights the peculiarity of these masers being so rare. To explain the rarity of
the \formaldehyde{} masers, \citet{Hoffman2003} suggested that the inversion of the
\gstate{} transition is due to a rare collisional excitation and not to excitation by a
free-free radiation field as proposed by \citet{Boland1981}. \citet{vanderwalt2022} and
\citet{vanderwalt2024} revisited the problem of the pumping of the 4.8 GHz \formaldehyde{}
masers and evaluated the critique raised against the \citet{Boland1981}
model. \citet{vanderwalt2022} and \citet{vanderwalt2024} proposed a pumping scheme that
explains the inversion of the \gstate{} transition and showed that the 4.8 GHz masers can
be pumped collisionally as well as by a free-free radiation field. \citet{vanderwalt2024}
argued that the radiative pumping via a free-free radiation field will be the dominant
mechanism in the environment of young high-mass stars. Still, no clear explanation for the
rarity of the masers has been proposed.

Another outstanding question regards the non-detection of 14.5 GHz \formaldehyde{}
masers. Although \citet{vanderwalt2024} demonstrated, in support of \citet{Boland1981},
that a free-free radiation field effectively inverts the \gstate{} transition, their
results also suggest that the \fstate{} transition (14.5 GHz) is typically inverted under
similar conditions. However, no 14.5 GHz maser emission associated with known 4.8 GHz
masers has yet been detected \citep{Hoffman2003,Chen2017a,shuvo2021}. Given that the 4.8
GHz masers are exclusively associated with high-mass star-forming regions
\citep{Araya2015} and that free-free radiation fields can invert both the \gstate{} and
\fstate{} transitions, the non-detection of 14.5 GHz masers requires explanation within
the framework of free-free pumping. One proposed explanation is that the 4.8 GHz masers
are projected offset from the centre of the \ion{H}{II} region, where the 14.5 GHz
background brightness temperature is diminished \citep{vanderwalt2024}, which will be
further addressed in this paper.

The results presented by \citet{vanderwalt2024} suggest that the \gstate{} and
\fstate{} transitions are inverted under the same conditions in the \nhtwo{}-\tk{} plane.
However, closer consideration of this way of presenting the behaviour of these two
transitions shows that it is not fully representative of the circumstellar environment,
where radial gradients in density, kinetic temperature, the dilution factor, and possibly
also the abundance of \formaldehyde{} exist. Pilot calculations of the inversion of the
\gstate{} and \fstate{} transitions as a function of distance from the outer radius of the
\ion{H}{II} region showed that there are combinations of \nhtwo{}, \tk{}, and dilution
factor where only the \gstate{} transition is inverted. Representative results will be
presented as a possible explanation for the non-detection of the 14.5 GHz masers.

Another factor that may influence the statistics of both the 4.8 and 14.5 GHz masers is
the attenuation of these masers by \formaldehyde{} absorption as the maser emission
propagates through the outer parts of the molecular envelope. Estimates of optical depths
at 4.8 and 14.5 GHz are presented, from which it follows that attenuation of the maser
emission can be significant and may be a contributing factor to the rarity of the 4.8 GHz
masers. 

\section{Numerical calculations}

The photo-ionisation code Cloudy (version 23.01; \citealt{Chatzikos2023}) was utilised to
simulate spherical \ion{H}{II} regions and compute the emitted free-free spectral energy
distribution (SED) integrated over the surface of the \ion{H}{II} region. This approach
enabled the consideration of free-free SEDs with non-constant electron density and/or
electron temperature throughout the ionised gas volume. The \ion{H}{II} region was
modelled as an ionised shell bounded by an inner radius, $r_i$, and an outer radius, $r_o$
\citep{Avalos2006,Lizano2008}. Temperatures and radii for luminosity class V ionising
stars of various spectral types were taken from \citet{Sternberg2003}.
In realistic \ion{H}{II} regions, $r_i$ is determined by the balance between the ram
pressure of the stellar wind and the thermal pressure at the inner boundary, as well as
the evolutionary state of the system. For the present calculations, $r_i$ was treated as a
free parameter. Except if stated otherwise, the outer radius, $r_o$, was set to $2.61
\times 10^{16}$ cm, the average radius of 12 hyper-compact \ion{H}{II} regions analysed by
\citet{Yang2021}, consistent with the hyper-compact phase. The radial density of
\ion{H}{I} is described by
\begin{equation}
  n(r) = n_0(r_i) \left(1 + \frac{\Delta r}{R_s}\right)^{-\alpha} ~\mathrm{cm^{-3}},
  \label{eq:nr}
\end{equation}
where $n_0(r_i)$ is the density at the inner radius, $\Delta r$ is the distance into the
\ion{H}{II} region from $r_i$, and $R_s$ is the scale length. When $R_s = r_i$, the
expression simplifies to $n(r) = n_0(r_i)(r/r_i)^{-\alpha}$. The chemical abundances
provided with Cloudy for \ion{H}{II} regions were adopted. The details of solving for the
level populations from the rate equations are given in \citet{vanderwalt2022} and
\citet{vanderwalt2024}.

\section{Results}
\subsection{Non-inversion of the \fstate{} transition}   
\label{section:noninversion}
\begin{figure}[!th]
   \centering
   \includegraphics[width=0.9\columnwidth]{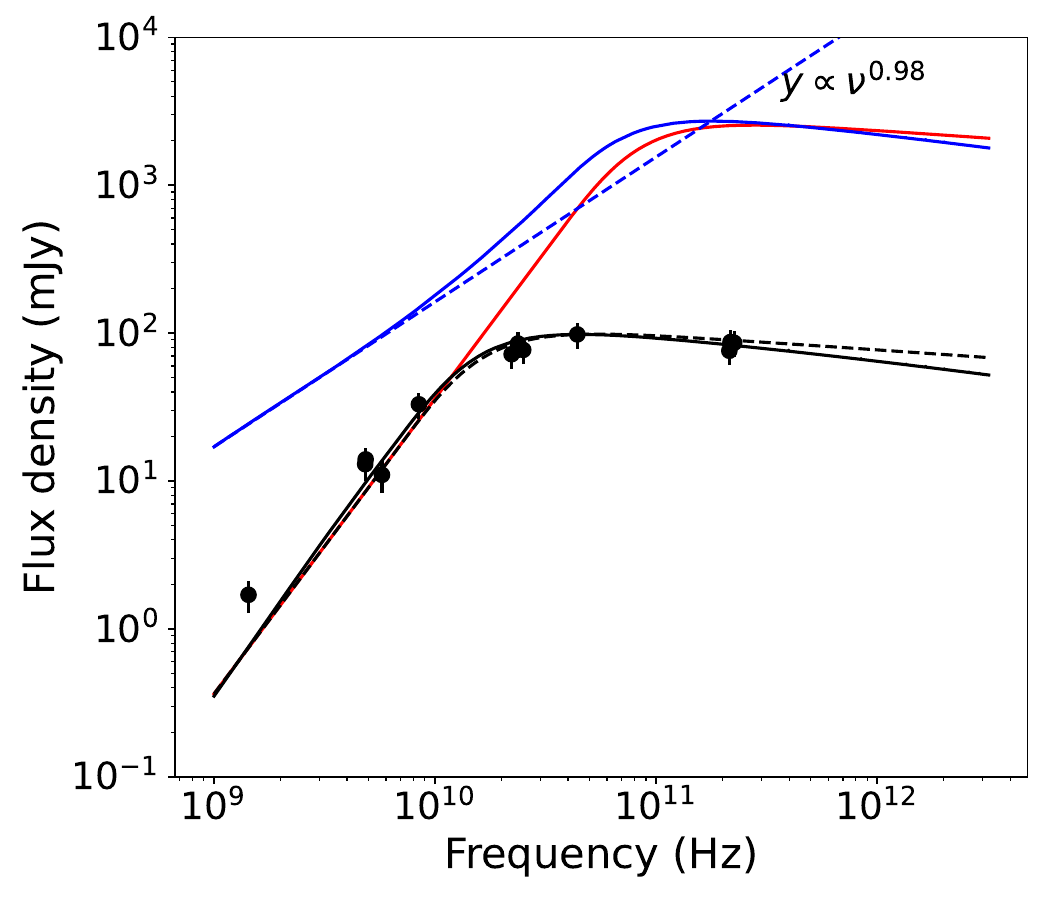}
      \caption{Comparison of the free-free SEDs used as
        examples. Filled black circles represent data for G24.78+0.08 A1 \citep{Cesaroni2019}. The solid
        black line is the SED calculated with Cloudy and normalised to the data at 44 GHz. The dashed black line is the theoretical SED with $T_K = 10^4$K and EM = $1.3 \times
        10^9\,\mathrm{pc\,cm^{-6}}$.\ The solid blue line is the SED calculated with Cloudy for the
        second example. The solid red line is a theoretical SED with $T_K = 10^4$K and EM = $3.9 \times
        10^{10}\,\mathrm{pc\,cm^{-6}}$.}
         \label{fig:g24spec}
\end{figure}

As a first example, Cloudy was used to model the free-free SED of the hyper-compact
\ion{H}{II} region G24.78+0.08 A1. \citet{Cesaroni2019} modelled the observed SED as
originating from a thin shell with an inner radius of 912 AU ($1.37 \times 10^{16}$ cm),
an outer radius of 1025 AU ($1.54 \times 10^{16}$ cm), a constant electron density of $n_e
= 6.9 \times 10^5~\mathrm{cm^{-3}}$, and an O9.5V ionising star. Employing these
parameters in a Cloudy simulation, a satisfactory fit to the observed flux density was
obtained with $n_e = 1.6 \times 10^6~\mathrm{cm^{-3}}$, as depicted by the solid black
line in Fig.\,\ref{fig:g24spec}. The SED, computed using Cloudy, was normalised to the
observed flux density at 44 GHz. The dashed black line represents the SED for a
theoretical \ion{H}{II} region with an electron temperature of $T_e = 10^4$ K and an
emission measure of $1.3 \times 10^9~\mathrm{pc\,cm^{-6}}$.

\begin{figure}[!ht]
   \centering
   \includegraphics[width=0.9\columnwidth]{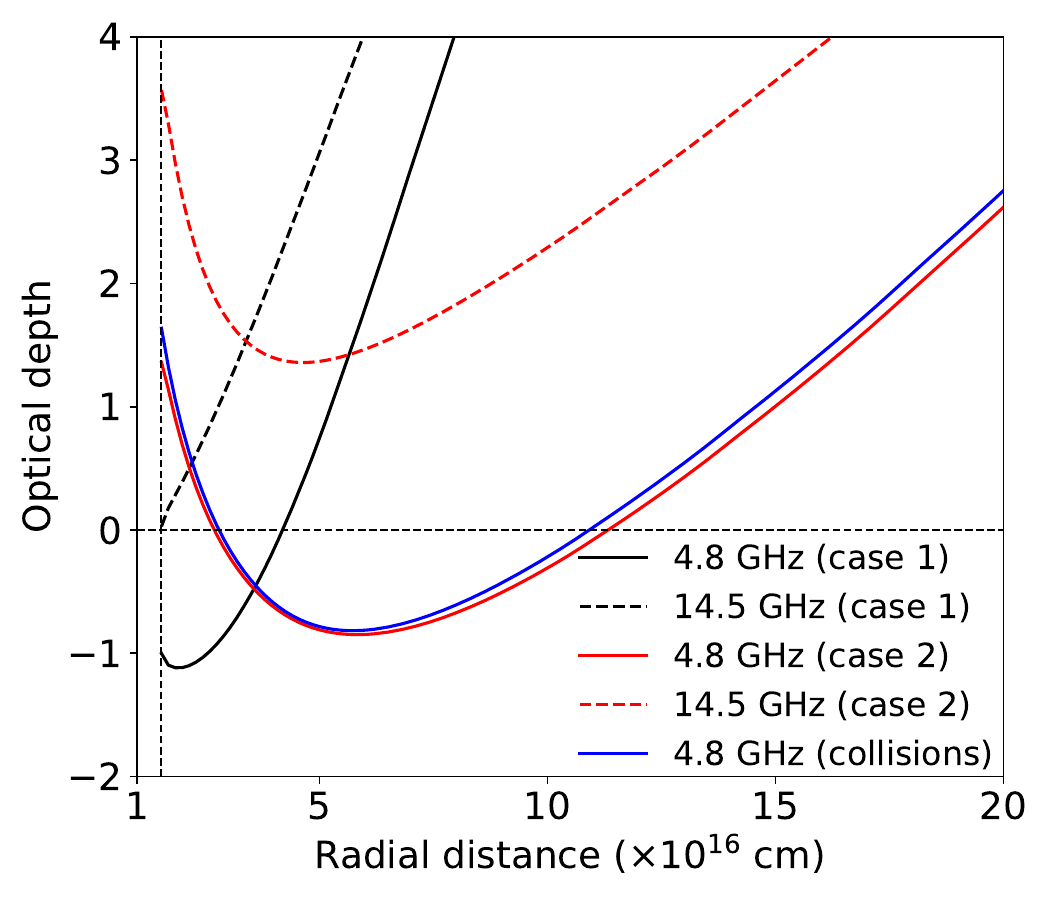}
      \caption{Variation in \taug{} and \tauf{} as a function of radial distance from the
        ionising star when the Cloudy SED for the hyper-compact \ion{H}{II} region
        G24.78+0.08 A1 is used. The vertical dashed line represents the outer radius of
        the \ion{H}{II} region. For case 1, $n_{H_2}(r_o) = 10^5~\mathrm{cm^{-3}}$ and
        $T_K(r_o) = 150$ K. For case 2, $n_{H_2}(r_o) = 7\times10^5~\mathrm{cm^{-3}}$ and
        $T_K(r_o) = 300$ K }
         \label{fig:tauvsrg24}
\end{figure}

Using the Cloudy SED as the pumping radiation field, optical depths \taug{} and \tauf{} as
a function of distance into the molecular cloud were computed, without beaming, adopting
$n_{\mathrm{H_2}}(r) = n_{\mathrm{H_2}}(r_o)(r/r_o)^{-p}$ and $T_{\mathrm{K}}(r) =
T_{\mathrm{K}}(r_o)(r/r_o)^{-q}$ with $p = 1.5$ and $q = 0.4$ \citep{vandertak2000} for
two cases: (1) $n_{\mathrm{H_2}}(r_o) = 10^5~\mathrm{cm^{-3}}$, $T_{\mathrm{K}}(r_o) =
150~\mathrm{K}$; (2) $n_{\mathrm{H_2}}(r_o) = 7 \times 10^5~\mathrm{cm^{-3}}$,
$T_{\mathrm{K}}(r_o) = 300~\mathrm{K}$. The \formaldehyde{} abundance was assumed constant
at $5.0 \times 10^{-6}$. The results for these two cases are presented in
Fig.\,\ref{fig:tauvsrg24}. The \gstate{} transition is inverted over a smaller radial
distance range for case 1 than for case 2, whereas the \fstate{} transition remains
non-inverted in both cases. The inversion of the \gstate{} transition, as depicted in
Fig.\,\ref{fig:tauvsrg24}, is, however, dominated by collisional rather than radiative
pumping. The solid blue line in Fig.\,\ref{fig:tauvsrg24} illustrates the variation in
\taug{} for case 2 when the free-free radiation field is absent, indicating a negligible
contribution from the radiation field to the inversion of the \gstate{} transition. The
effect of beaming was found to have minimal impact on \taug{}. For case 2, inversion of
the \gstate{} transition occurs for $3.5 \times 10^4 < n_{\mathrm{H_2}} < 3.2 \times
10^5~\mathrm{cm^{-3}}$ and $90 < T_{\mathrm{K}} < 160~\mathrm{K}$.

\begin{figure}[!ht]
   \centering \includegraphics[width=0.9\columnwidth]{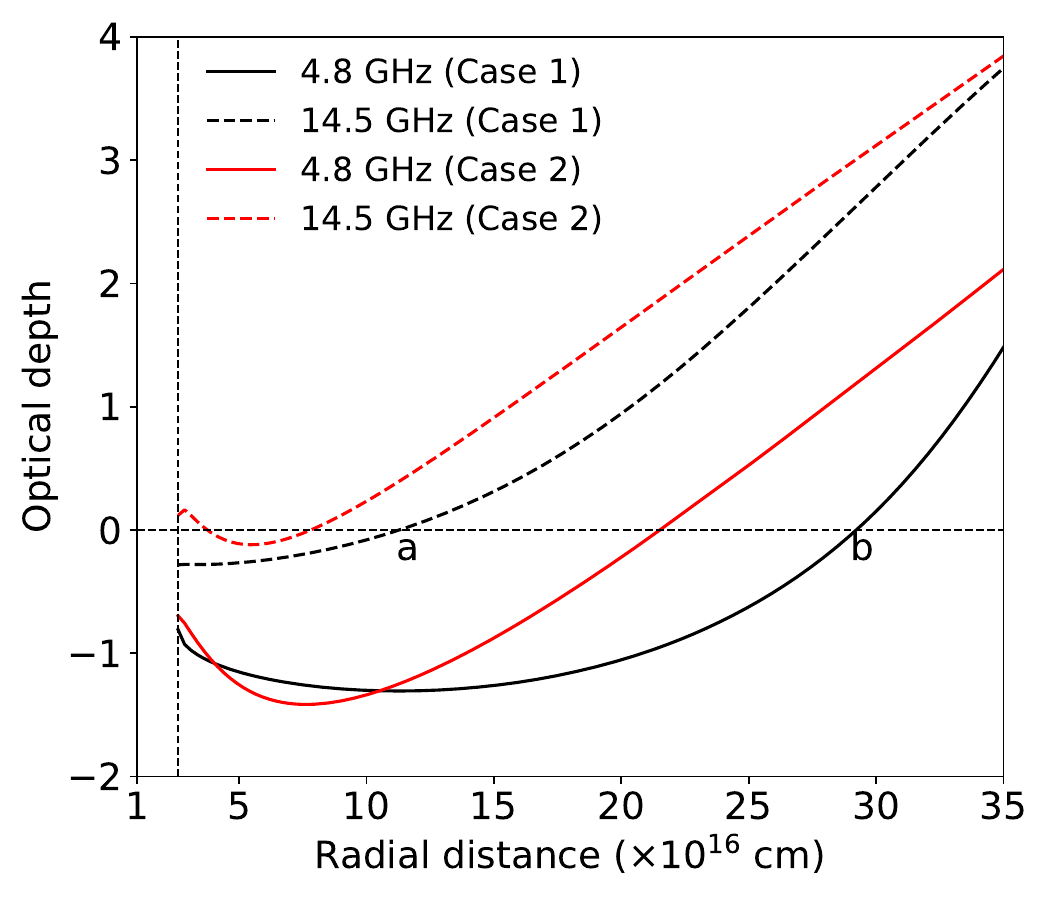}
      \caption{Variation in \taug{} and \tauf{} as a function of radial distance from the
        ionising star when the Cloudy SED for the second example (blue line in
        Fig.\,\ref{fig:g24spec}) is used. Cases 1 and 2 correspond to those used in
        Fig.\,\ref{fig:tauvsrg24} }
         \label{fig:tauvsr3}
\end{figure}

Second, an \ion{H}{II} region with an inner radius of $r_i = 1.58 \times
10^{15}~\mathrm{cm}$ and an outer radius of $r_o = 2.61 \times 10^{16}~\mathrm{cm}$ was
considered. The initial \ion{H}{I} radial distribution is described by Eq.\,\ref{eq:nr},
with a density at the inner radius of $n_0(r_i) = 3.16 \times 10^7~\mathrm{cm^{-3}}$, a
scale length of $R_s = 3.2 \times 10^{14}~\mathrm{cm}$, and $\alpha = 2$. These parameters
were selected to produce a free-free SED that is flatter than $\nu^2$ at frequencies
below the turnover frequency. The resulting SED, depicted as the solid blue line in
Fig.\,\ref{fig:g24spec}, was scaled using the same normalisation as for G24.78+0.08 A1. The
lower-frequency portion of the SED exhibits a $\nu^{0.98}$ dependence, as indicated by the
dashed blue line. The solid red line represents the theoretical SED for an \ion{H}{II}
region with an electron temperature of $T_e = 10^4~\mathrm{K}$ and an emission measure of
$3.9 \times 10^{10}~\mathrm{pc\,cm^{-6}}$. The corresponding variation in \taug{} and
\tauf{} with distance is shown in Fig.\,\ref{fig:tauvsr3}.

Comparison of Fig.\,\ref{fig:tauvsr3} with Fig.\,\ref{fig:tauvsrg24} shows that the
variation in \taug{} and \tauf{} deviates significantly from the case of
collision-dominated inversion. The radial distances at which the \gstate{} transition
switches from inversion to non-inversion are shifted to larger values, indicating the
effect of the radiation field. The important point to note is that, in both cases, a
distance interval exists where only the \gstate{} transition is inverted. In case 1, the
\fstate{} transition is not inverted between points $a$ and $b$, where point $a$
corresponds to $R = 1.12 \times 10^{17}~\mathrm{cm}$, $n_{\mathrm{H_2}} = 1.1 \times
10^4~\mathrm{cm^{-3}}$, $T_{\mathrm{K}} = 83~\mathrm{K}$, and $W = 0.014$, and point $b$
corresponds to $R = 2.9 \times 10^{17}~\mathrm{cm}$, $n_{\mathrm{H_2}} =
2680~\mathrm{cm^{-3}}$, $T_{\mathrm{K}} = 57~\mathrm{K}$, and $W = 0.002$.

\begin{figure}[!ht]
   \centering
   \includegraphics[width=0.9\columnwidth]{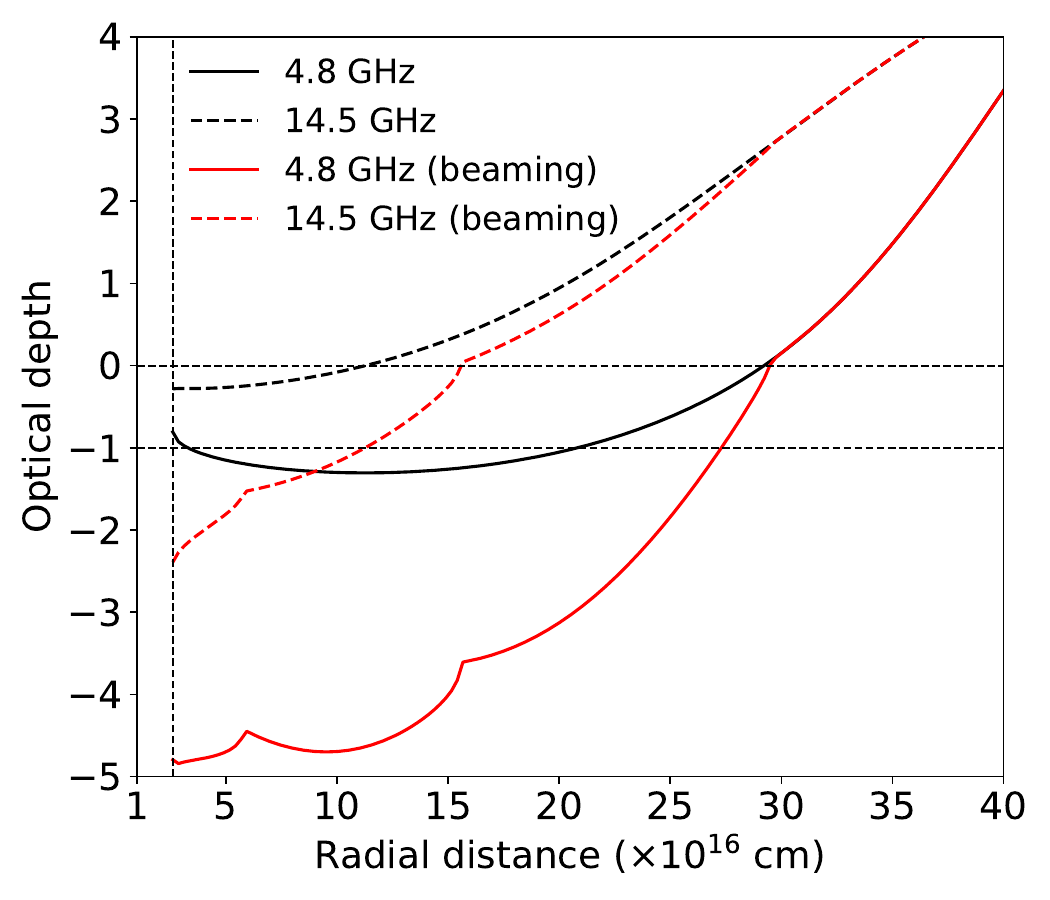}
   \caption{Variation in \taug{} as a function of radial distance from the ionising star
     when the Cloudy SED for the second example (blue line in Fig.\,\ref{fig:g24spec}) is
     used and when beaming is included.}
     \label{fig:tauvsrbeaming}
\end{figure}

Figure \ref{fig:tauvsrbeaming} compares the variation in \taug{} with and without beaming
for case 1, employing Eq.\,8 of \citet{vanderwalt2024} with $\alpha = 5$ (not to be
confused with $\alpha$ in Eq.\,\ref{eq:nr}). A range of conditions persists where the
\gstate{} transition is inverted, whilst the \fstate{} transition remains non-inverted.

\subsection{Masers are projected offset from the centre of the \ion{H}{II} region}
\label{section:projection}

To account for the non-detection of 14.5 GHz masers, \citet{vanderwalt2024} proposed that
the masing region is projected towards the edge of a hyper-compact \ion{H}{II} region,
where the 14.5 GHz brightness temperature is diminished.  Among the known 4.8 GHz maser
sources, it is only NGC 7538 IRS1 for which high-resolution C and U-band maps as well as
accurate positions of the 4.8 GHz masers could be found to test the proposal of
\citet{vanderwalt2024}. Considering Fig. 2 of \citet{Sandell2009} and the coordinates for
the 4.8 GHz masers given by \citet{Hoffman2003}, it is found that the offset in right
ascension between the masers and the reference position is $-0.42$ arc-seconds and $-0.064$
arc-seconds in declination. Inspection of Fig. 2 of \citet{Sandell2009} shows that, while
there is 4.8 GHz background emission at the position of the masers, there is either very
weak or no background continuum emission at 14.96 GHz at the same position. Further
inspection of the 1.2 cm brightness temperature map of NGC 7538 IRS1 presented by
\citet{Beuther2017} shows a definite absence of 1.2 cm emission at the position of the 4.8
GHz masers, in agreement with the K-band image in Fig. 2 of \citet{Sandell2009}.

   \begin{figure}[h!]
   \centering
   \includegraphics[width=0.9\columnwidth]{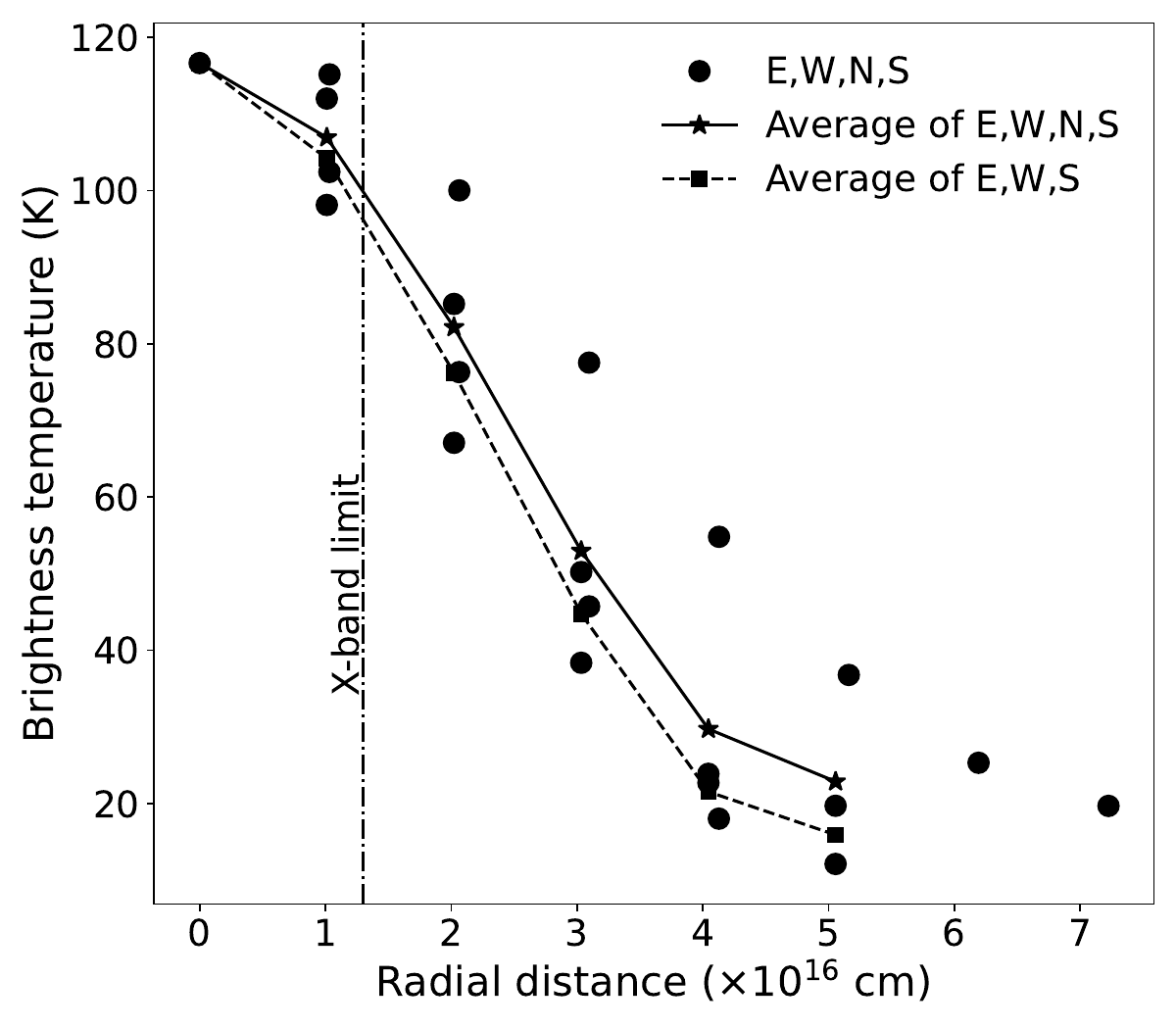}
      \caption{Variation in the free-free brightness temperature for G32.7441-0.0755 as
        derived from the publicly available CORNISH data. The vertical line represents the
      outer edge of the X-band emission estimated from the image in \citet{Yang2021}.}
         \label{fig:g32}
   \end{figure}

Although an accurate position of the 4.8 GHz maser in G32.7441-0.0755 is not available,
the high-resolution 5 GHz map from the CORNISH survey \citep{Hoare2012,Purcell2013} could
be used to assess the variation in free-free brightness temperature with projected radial
distance from the emission centre. The free-free emission exhibits asymmetry, with
elongation towards the north. The east-west extent of the free-free emission, derived from
the CORNISH survey image, is approximately 1.29 arc-seconds. Based on a local standard of
rest velocity ($V_\mathrm{lsr}$) of 36.4 $\mathrm{km\,s^{-1}}$, the near- and far kinematic
distances are calculated as 2.36 kpc and 11.73 kpc, respectively.  At the far kinematic
distance, this corresponds to a linear dimension of 0.74 pc, which exceeds the typical
size of a hyper-compact \ion{H}{II} region. Consequently, the near kinematic distance was
adopted. Figure \ref{fig:g32} presents the brightness temperature as a function of
projected radial distance, which shows a pronounced decrease with increasing radial
distance from the centre. From the X-band image of G32.7441-0.0755 \citep{Yang2021}, an
angular diameter of approximately 0.744 arc-seconds is found, corresponding to a projected
radial distance to the edge of the X-band emission of approximately $1.3 \times
10^{16}~\mathrm{cm}$, as indicated by the vertical dash-dot line in
Fig.\,\ref{fig:g32}. Consequently, a 14.5 GHz maser would likely exhibit negligible or no
emission if the 4.8 GHz maser is projected against the \ion{H}{II} region beyond the edge
of the X-band emission.

   \begin{figure}[h!]
   \centering
   \includegraphics[width=0.9\columnwidth]{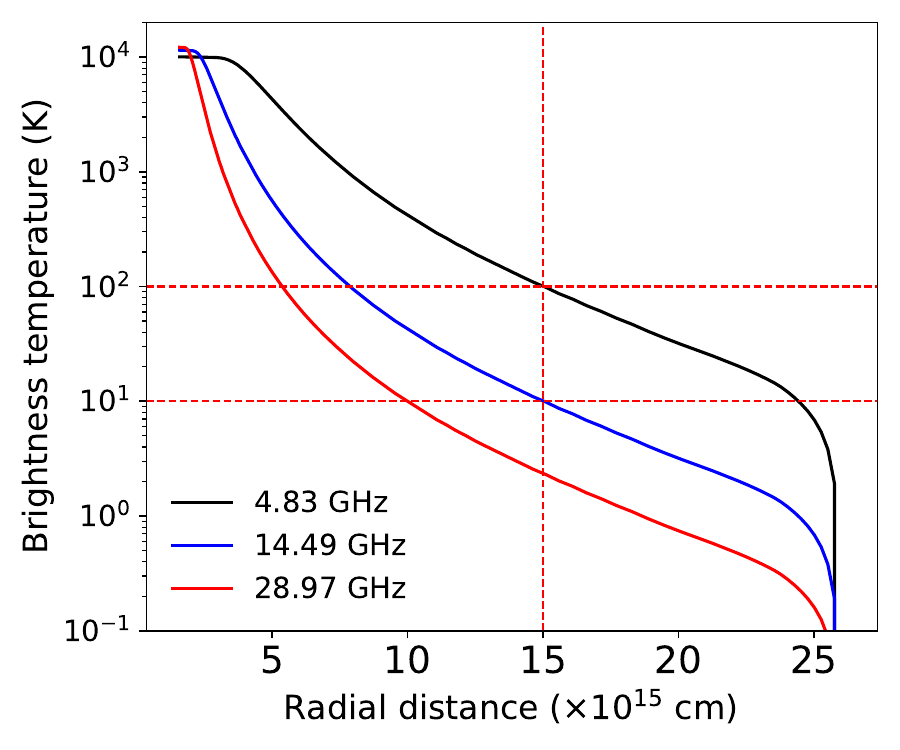}
      \caption{Radial distance dependence of the free-free brightness temperature at 4.8,
        14.5, and 28.9 GHz for the second example presented in
        Sect.\,\ref{section:noninversion}.  }
         \label{fig:rdist}
   \end{figure}

Figure\,\ref{fig:rdist} shows an example of the projected radial variation in the
continuum brightness temperatures at 4.8, 14.5, and 28.9 GHz calculated from the output of
the Cloudy simulation, for which the free-free SED is the solid blue line in
Fig.\,\ref{fig:g24spec}. It is seen that the brightness temperature at 4.8 GHz is
approximately an order of magnitude greater than at 14.5 GHz at distances around $1.5
\times 10^{16}$ cm. If, for example, in the case of G32.7441-0.0755 the masing column is
projected against the \ion{H}{II} region at a distance of $\sim 2 \times 10^{16}$ cm from
the centre, the background brightness temperature at 4.8 GHz is $\sim 60$ K. Assuming
then, from Fig.\,\ref{fig:rdist}, that the background brightness temperature at 14.5 GHz
is 6 K, \tauf{} = $-1$ and \taug{} = $-4.64$ (from Fig.\,\ref{fig:tauvsrbeaming}), the
brightness temperature of the 4.8 GHz maser is 6212 K and 16.3 K for the 14.5 GHz maser.
To what extent this scenario can explain the non-detection of the 14.5 GHz masers,
at least for some of the known 4.8 GHz maser sources, can only be evaluated with the aid
of high-resolution mapping of the background continuum emission at 4.8 and 14.5 GHz, as
well as having accurate positions of the 4.8 GHz masers, which is currently not available
for most of the 4.8 GHz masers.

\subsection{Attenuation of the 4.8 and 14.5 GHz maser emission}
\label{section:attenuate}
Observations by \citet{Okoh2014} using the Nanshan 25 m telescope indicate a significant
detection rate of \formaldehyde{} 4.8 GHz absorption towards high-mass star-forming
regions associated with 6.7 GHz \methanol{} masers. Five \formaldehyde{} maser
sources—G23.01-0.41, G23.71-0.20, G24.33+0.13, G25.83-0.18, and G37.55+0.20—are included
in the sample and exhibit associated 4.8 GHz absorption features. At 4.8 GHz, the beam
size of the Nanshan 25-m telescope is 10 arc-minutes, corresponding to a diameter of 8.7 pc
at a distance of 3 kpc. Although this exceeds the scale of hyper-compact and ultra-compact
\ion{H}{II} regions, it underscores the necessity of considering attenuation of
\formaldehyde{} 4.8 and 14.5 GHz maser emission within the circumstellar envelope.

The optical depth at the line centre along a ray path through the molecular
envelope is expressed, in terms of the level populations, as
\begin{equation}
  \tau_\nu = 
  \frac{1}{8\pi}\left(\frac{c}{\nu}\right)^3\frac{1}{\Delta\varv}\frac{g_u}{g_\ell}A_{u\ell}\int_{r_{min}}^{r_{max}}
  n_\ell(r)[1 - \frac{g_\ell n_u(r)}{g_un_\ell(r)}]dr  \label{eq:tau1}
\end{equation}
or, in terms of the excitation temperature for the transition, as
\begin{equation}
\tau_\nu = \frac{1}{8\pi}\left(\frac{c}{\nu}\right)^3\frac{1}{\Delta\varv}\frac{g_u}{g_\ell}A_{u\ell}\int_{r_{min}}^{r_{max}}
       n_\ell(r)[1 - \exp(\frac{-h\nu}{kT_{ex}(r)})]dr
       \label{eq:tau2}
,\end{equation}
where all symbols have their normal meaning and a constant velocity width, $\Delta \varv$,
has been assumed.

Due to the overpopulation of the lower doublet states relative to local thermodynamic
equilibrium under specific conditions \citep{Townes1969,Evans1975,Troscompt2009},
optical depths cannot be estimated assuming  local thermodynamic
equilibrium level populations, as performed, for
instance, by \citet{vanderwalt2021} for CS. Theoretical estimates of the optical depths
must therefore be made by solving for the level populations from the rate equations
without a free-free radiation field and assuming radial dependences for \nhtwo{}, kinetic
temperature, \tk{}, and the \formaldehyde{} abundance, \xhtwoco{}. For the following
calculations $n_{\mathrm{H_2}}(r) =
n_{\mathrm{H_2}}(r_{\mathrm{min}})(r/r_{\mathrm{min}})^{-1.5}$ and $T_{\mathrm{K}}(r) =
T_{\mathrm{K}}(r_{\mathrm{min}})(r/r_{\mathrm{min}})^{-0.4}$ \citep{vandertak2000,
  Gieser2021} were used with $n_{\mathrm{H_2}}(r_{\mathrm{min}}) = 3\times
10^5~\mathrm{cm^{-3}}$ and $T_{\mathrm{K}}(r_{\mathrm{min}}) = 100~\mathrm{K}$. The inner
radius was set to $r_{\mathrm{min}} = 2.9 \times 10^{17}~\mathrm{cm}$, where inversion of
the \gstate{} transition ceases in Figs.\,\ref{fig:tauvsr3} and
\ref{fig:tauvsrbeaming}. The outer radius was set to $r_{\mathrm{max}} = 1$ parsec.

\begin{figure}[!ht]
   \centering
   \includegraphics[width=0.9\columnwidth]{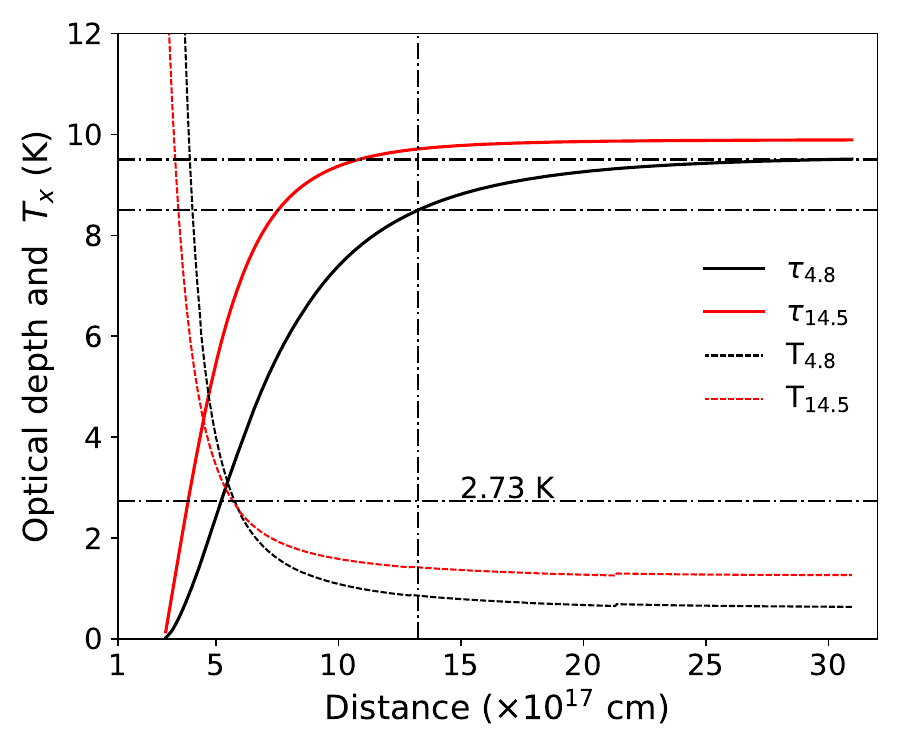}
      \caption{Cumulative optical depth and excitation temperatures as a function of
        distance at 4.8 GHz and 14.5 GHz.}
         \label{fig:tauplot1}
\end{figure}

Level populations were computed as a function of radial distance in intervals of $5 \times
10^{15}~\mathrm{cm}$ for three cases of $X_{\mathrm{H_2CO}}(r)$, with
$X_{\mathrm{H_2CO}}(r_{\mathrm{min}}) = 10^{-6}$ based on \citet{Nomura2004}. In the first
case, $X_{\mathrm{H_2CO}}(r) \propto r^{-\gamma}$ was assumed, with
$X_{\mathrm{H_2CO}}(r_{\mathrm{max}}) = 10^{-9}$, implying $\gamma = 2.9$. The cumulative
optical depths and excitation temperatures at 4.8 and 14.5 GHz as a function of radial
distance are presented in Fig.\,\ref{fig:tauplot1}. The excitation temperatures decrease
rapidly, falling below 2.73 K. The cumulative optical depth increases with distance into
the envelope and stabilises as \nhtwo{} and \xhtwoco{} decrease. At 1 parsec, the optical
depths are \taug{} = 9.5 and \tauf{} = 9.9. Thus, for a maser for which, say, \taug{}
$\sim -9$ and which is projected against a free-free radiation field with a brightness
temperature of $T_b = 100~\mathrm{K}$, the unattenuated brightness temperature is
approximately $8 \times 10^5~\mathrm{K}$. With attenuation in the circumstellar envelope,
the brightness temperature is reduced to approximately 60 K.

In a second case, the radial abundance profile of \formaldehyde{} from Fig.\,4(b) of
\citet{Nomura2004} was used, where \xhtwoco{} = $10^{-6}$ up to approximately 0.3 pc,
followed by a sharp decline to $10^{-9}$. The cumulative optical depths exhibit a
behaviour similar to that in Fig.\,\ref{fig:tauplot1}, with \taug{} and \tauf{} reaching
maximum values of approximately 29 and 32, respectively. Under these conditions, 4.8 and
14.5 GHz maser emission would be entirely absorbed in the envelope.  \citet{Nomura2004}
also consider a scenario where mantle molecules are trapped in water ice, therefore
evaporating at smaller radial distances. In this case, \xhtwoco{} decreases sharply from
$10^{-6}$ to $10^{-9}$ at approximately 0.09 pc (see their Fig.\,5b). Due to this rapid
decrease in \xhtwoco{} near $r_{\mathrm{min}}$, the optical depths are \taug{} $\approx
0.16$ and \tauf{} $\approx 0.1$ at 1 parsec, which means that the maser radiation can
escape unattenuated.

\section{Discussion and conclusions}

As stated in the Introduction, the reasons behind the rarity of the \formaldehyde{} masers
remain unknown. The result that the free-free SED for G24.78+0.08 A1 is ineffective in
inverting the \gstate{} transition is important in the context of the rarity of the 4.8
GHz \formaldehyde{} masers. If a significant number of young high-mass star-forming
regions have free-free SEDs similar to that of G24.78+0.08 A1, it follows that, most
likely, none of these will have associated 4.8 GHz masers. Thus, part of the answer as to
why the 4.8 GHz \formaldehyde{} masers are rare may therefore be that the free-free
radiation fields of some high-mass star-forming regions are not effective to invert the
\gstate{} transition. A comparison of the free-free SEDs of high-mass star-forming regions
with and without 4.8 GHz masers may be helpful in this regard.

 Although free-free SEDs similar to that of G24.78+0.08 A1 may not suffice to invert
 the \gstate{} transition, collisional inversion may still be possible, as suggested in
 Fig.\,\ref{fig:g24spec} and discussed by \citet{vanderwalt2022}. However, for purely
 collisional pumping, \taug{} is typically less than 1 and, as noted above, beaming does
 not have a significant effect on the inversion. It is therefore expected that, considering the
 possibility of attenuation in the outer regions of the circumstellar envelope, collisional
 inversion will give rise to weak masers that might not be detected.

The results shown in Figs.\,\ref{fig:tauvsr3} and \ref{fig:tauvsrbeaming} are significant
since they suggest the possibility that there are regions in the circumstellar envelope
where only the \gstate{} transition is inverted. Detection of 14.5 GHz masers associated
with the 4.8 GHz masers is therefore not a necessary condition for the radiative pumping
by a free-free radiation field. However, the non-detection of a 14.5 GHz maser does not
imply that the 4.8 GHz maser is excited in a region where the \fstate{} transition is not
inverted. It is still possible that \tauf{} is small and the masing region is projected
against some part of the \ion{H}{II} region where the 14.5 GHz background free-free
emission is weak, or absent, as discussed in Sect.\,\ref{section:projection}.

Considering the results in Sect.\,\ref{section:attenuate}, the question is to what degree
the 4.8 GHz (and perhaps the 14.5 GHz) \formaldehyde{} masers are attenuated in the
molecular envelope. \citet{Gong2023} present maps of \taug{} for Cygnus X, which
show that \taug{} is $\lesssim$ 0.5 for extended cloud structures that are well resolved at
an angular resolution of $10^\prime.8$. It can be assumed that similar extended regions
of \formaldehyde{} absorption are associated with high-mass star-forming regions in
general. However, attenuation for \taug{}$\lesssim$ 0.5 is not enough to affect the maser
emission significantly.  

Although the excitation temperatures are plotted against distance in
Fig.\,\ref{fig:tauplot1}, the graph implicitly reflects how the excitation temperatures
vary with density and temperature. The excitation temperatures drop below 2.73 K at
\nhtwo{}$\sim 1.08\times 10^5~\mathrm{cm^{-3}}$ and \tk{}$\sim$ 76 K and stay below 2.73 K
up to 1 parsec, where \nhtwo{}$\sim 8.6 \times 10^3\,\mathrm{cm^{-3}}$ and \tk{}$\sim$ 38
K. This range of densities and kinetic temperatures fall within the range of densities and
kinetic temperatures derived for high-mass star-forming regions \citep[see
  e.g.][]{vandertak2000,McCauley2011,Ginsburg2011}. The implication is that a significant
fraction of \formaldehyde{} molecules might be in \zerostate{} over a large part of the
`inner' envelope, and that some degree of attenuation of maser emission propagating
through the envelope is therefore inevitable.

Further evidence that the optical depths as shown in Fig.\,\ref{fig:tauplot1} are not
unrealistic is found in \citet{Gieser2021}, who derived an average \formaldehyde{} column
density of $\sim 4 \times 10^{15}\,\mathrm{cm^{-2}}$ for 18 high-mass star-forming
regions. Assuming an excitation temperature of 1.5 K for the \gstate{} transition and
$\Delta \varv = 5\,\mathrm{km\,s^{-1}}$ in Eq.\,\ref{eq:tau2}, one finds that $\tau = 6.89
\times 10^{-14}N_\ell$, where $N_\ell$ is the column density of molecules in
\zerostate{}. Using an ortho-to-para ratio of 3:1 and assuming that 50\% of the molecules are in
\zerostate{}, giving $N_\ell = 1.5\times 10^{15}\,\mathrm{cm^{-2}}$, an optical depth of
14.6 is found. A more relevant example is that of NGC 7538 IRS1, where \citet{Feng2016}
estimated a \formaldehyde{} column density of approximately $10^{17}\,\mathrm{cm^{-2}}$
close to the position of the 4.8 GHz \formaldehyde{} maser. If the excitation temperature
of the \gstate{} transition is $\approx$ 1.5 K for only 10\% of the total column density
of o-\formaldehyde{}, an optical depth of $\sim$25 is found.  It is also to be noted that
the p-\formaldehyde{} lines at 218.33 GHz ($3_{03}-2_{02}$), 218.475 GHz
($3_{22}-2_{21}$), and 218.76 GHz ($3_{21}-2_{20}$), as well as the
$\mathrm{C^{18}O(2-1)}$ (219.56 GHz) and $\mathrm{^{13}CO(2-1)}$ (220.398 GHz) lines, are
in absorption against the background free-free emission, signifying the presence of dense
gas in front of the 1.37 mm continuum source. \citet{Goddi2015} independently concluded
that NGC 7538 IRS1 is the densest known hot core, with the implication that the location of
the masing \formaldehyde{} region must be such that the `outward' propagating maser
emission can escape with minimum attenuation.  It might therefore be that the 22 known 4.8
GHz \formaldehyde{} masers are indicative of a special or rare geometry that allows the
maser emission to escape with minimum attenuation.

 In this regard, it is worth noting the following: of the 22 known 4.8 GHz masers, only in the case of Sgr B2N A does the maser line lie in an absorption feature
 \citep{Mehringer1994}. Other maser sources for which an absorption feature is present in
 the maser spectrum (depending on the angular resolution or the angular size of the region
 of integration) are Sgr B2N C, Sgr B2S H \citep[][VLA-CnB]{Mehringer1994}, G29.96-0.02
 \citep[][VLA-CnB]{Hoffman2003}, G25.83-0.18 \citep[][VLA-BnA, VLA-D]{Araya2008},
 G32.74-0.07 \citep[][Arecibo]{Araya2015}, IRAS 18566+0408, NGC 7538 IRS1
 \citep[][GBT]{Araya2007d}, and G339.88-1.26 \citep[][TMRT]{Chen2017a}. In these cases, the
 4.8 GHz maser lines lie at the edge of the absorption feature. The velocities of the
 absorption features were found to coincide with the systemic velocities for those sources
 for which the systemic velocities could be found. The maser lines are typically offset
 from the systemic velocity by a few $\mathrm{km\,s^{-1}}$. No absorption features in the
 maser spectra have been reported for Sgr B2 B, Sgr B2S D, Sgr B2 E, Sgr B2N F, Sgr B2S G,
 Sgr B2 I \citep[][VLA-CnB]{Mehringer1994}, G0.38+0.04 \citep[][ATCA]{Ginsburg2015},
 G23.71-0.20 \citep[][VLA-A]{Araya2006}, G23.01-0.41 \citep[][VLA-BnA, VLA-D]{Araya2008},
or  G24.33+0.13 \citep[][ATCA]{McCarthy2022}. The absorption features referred to above are
 most likely against the cosmic microwave background \citep[see e.g.][]{Araya2015}.

It is remarkable that for several high-mass star-forming regions, the 4.8 GHz maser
spectra have the same characteristic, i.e. the maser emission lies at the edge of an
absorption feature and is therefore offset from the systemic velocity. In none of the
cases where a 4.8 GHz maser line lies at the edge of an absorption feature is there an
indication of maser emission in the velocity range covered by the absorption
feature. Using the centre velocity, $\varv_{abs}$, and the standard deviation,
$\sigma_{abs}$, of the absorption features listed by \citet{Okoh2014} and the published
velocities for the 4.8 GHz \formaldehyde{} masers in G23.01-0.41, G24.33+0.13, and
G25.83-0.18, the velocity offset relative to the velocity dispersion,
$|\varv_{maser}-\varv_{abs}|/\sigma_{abs}$, for these sources is respectively 0.62, 1.47, and 2.08. For G339.88-1.26, a value of 2.34 is found for the maser feature nearest
to the absorption feature and 3.77 for the strongest 4.8 GHz maser feature using the
results presented by \citet{Chen2017a}; for G29.96-0.20, a value of 1.54 using the results
of \citet{Hoffman2003}; and for G37.55+0.20 a value of 3.11 using the results of
\citet{Araya2004a}. Apart from G23.01-0.41, the maser velocity is offset from the centre
velocity of the 4.8 GHz absorption by more than one standard deviation of the velocity
dispersion.

A possible explanation for the observed offsets is that the \formaldehyde{} masers
originate in a kinematic structure that Doppler-shifts the maser emission to velocities
where the \taug{} in the envelope is sufficiently small, thereby allowing the maser emission
to escape. In this regard, it is noted that the observations of \citet[][see their
  Fig.7]{Beltran2011} suggest the presence of a rotating toroid in G29.96-0.02 with the
most redshifted emission at 100 $\mathrm{km\,s^{-1}}$. Components I and II of the 4.8 GHz
masers in G29.96-0.02 have velocities respectively of 100.24 and 102.0
$\mathrm{km\,s^{-1}}$. This suggests that the two maser components are physically
associated with the rotating toroidal structure, which can explain the offset of the maser
velocity from the systemic velocity.

As a last remark, it is noted that the results presented above do not apply to the
extragalactic formaldehyde masers. Whereas the Galactic \formaldehyde{} masers are
associated with individual high-mass star-forming regions, the extragalactic
\formaldehyde{} masers in IC 860, IRAS 15107+0724, and Arp 220 seem to arise from a
central molecular structure of size 30$-$100 pc centred on the nuclei of these galaxies as
well as from some isolated star-forming regions \citep{Baan2017}. The analysis and
interpretation of integrated maser emission over such large regions, which include many
star-forming regions, falls outside the scope of this paper. The same applies to the
single dish \gstate{} and \fstate{} absorption spectra of starburst galaxies presented by
\citet{Mangum2013}.

Based on the above-presented results, it is concluded that:
\begin{enumerate}
 \item Not all free-free SEDs associated with hyper-compact \ion{H}{II} regions are
   effective in inverting the \gstate{} transition. This may partially
   explain the small number of detected 4.8 GHz \formaldehyde{} masers. \label{conc:one}
\item The non-detection of the 14.5 GHz maser may partly be due to the masing region being
  projected against a region with very weak or no 14.5 GHz emission.
\item The numerical calculations suggest that there might be regions in the circumstellar
  environment where the \gstate{} transition is inverted, but not the \fstate{}
  transition. The non-detection of the 14.5 GHz \formaldehyde{} maser may be partly due to
  the inversion of the \gstate{} transition occurring under conditions where the \fstate{}
  transition is not inverted.
\item Attenuation of 4.8 GHz maser emission at velocities close to the centre
  velocity of \formaldehyde{} in the envelope can be significant and can even lead to
the  complete absorption of the maser emission, which can result in the non-detection of 4.8
  GHz masers. For at least six high-mass star-forming regions, the maser velocities
    are offset from the systemic velocity, which might result in a reduced attenuation of
    the masers. 
\item The small number of \formaldehyde{} masers in the Galaxy may be due to the combined
  effect of some free-free SEDs not being effective to invert the \gstate{} transition,
  attenuation of the maser emission in the molecular envelope, the chemistry of the
  star-forming region, and a short lifetime of the masers.
\end{enumerate}

\begin{acknowledgements}
  The author acknowledges financial support from the North-West University.
\end{acknowledgements}

\bibliographystyle{aa}
\bibliography{ref}

\end{document}